# Terahertz emission from metal nanoparticles due to ultrafast heating

DANIIL A. FADEEV, IVAN V. OLADYSHKIN,* VYACHESLAV A. MIRONOV

*Institute of Applied Physics of Russian Academy of Sciences, 46 Ul'yanov St. Nizhny Novgorod, Russia 603950*
*\*Corresponding author: oladyshkin@gmail.com*

**We demonstrate theoretically that ultrafast heating of metal nanoparticles by the laser pulse should lead to the generation of coherent terahertz (THz) radiation during the heat redistribution process. It is shown that after the femtosecond laser pulse action the time-dependent gradient of the electronic temperature induces low-frequency particle polarization with the characteristic timescale of about fractions of picosecond. In the case of the directed metallic pattern, the THz pulse waveform can be controlled by changing geometry of the individual particle. The generation mechanism proposed in this Letter can be used for interpretation of the recent experiments on the THz generation from metallic nanoparticles and nanostructures.**

http://dx.doi.org/10.1364/OL.99.099999

Conversion of femtosecond optical pulses to the THz radiation attracts a lot of attention not just because of a large number of THz waves' applications, but also due to fundamental interest to the fast nonlinear phenomena taking place in various materials. In particular, the processes of electrons' thermalization and recombination in conducting media like metals [1–3], semimetals [4], graphene [5–7] and topological insulators [8, 9] have the characteristic times in the range of 100 fs – 1 ps which corresponds to the frequency range of 1–10 THz. In recent years such phenomena have been actively investigated by optical pump – THz probe and THz pump – optical probe methods. In the material properties studies exploring of optical-to-THz conversion is worth it if the behind physical mechanisms are clear.

In the present paper we propose a mechanism of the laser-induced THz generation from metal particles or structured surfaces based on the thermal effects in the electron gas. By means of this work we would like to point out the fact that simple heating of electrons in a nanoparticle array may lead to coherent THz pulse generation and this effect should be taken into account. This paper generalizes the thermal model of THz generation from a flat metal surface [10] to the case of a structured metal. We found that in the array of nanoparticles one should expect significantly stronger temperature gradients than longitudinal gradients on the flat metal surface.

After the first experiments on the laser-induced THz generation from metal surfaces [11, 12] three main directions in the microscopic theory were developed. The first group of models is based on the plasmon excitation and acceleration of emitted electrons in the plasmon electric field [13, 14]. The second one involves generation of THz radiation due to ponderomotive force of the optical field [15, 16]. The third approach describes low-frequency field generation as a result of inhomogeneous heating and temperature dynamics of the electron gas near the surface [10, 17].

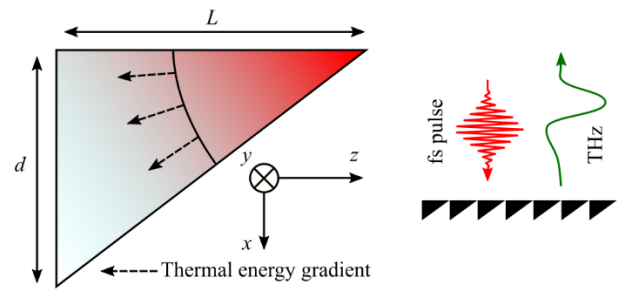

Fig. 1. Schematic picture of the particle and a thermal energy gradient inside (left); principal scheme of THz generation from the particle array (right).

In this Letter we will focus just on the thermal model of THz field generation [10, 17]. Ponderomotive-like mechanisms must be excluded because they describe instant quadratic response and give low-frequency radiation in the range of 10-30 THz (for 30-100 fs pump laser pulses) [15, 16]; at the same time the experimental spectral maximum is about 0.5-1 THz (see [11, 12] and later papers). Moreover, the full energy of the THz pulse in these models is strongly underestimated. Here we also do not consider widely discussed surface plasmon excitation and related processes [13, 14] and choose such a geometry that no plasmons can be produced by the laser pulse. Theoretical results presented in this Letter can be the basis for theoretical interpretation of the experiments [18–20] where laser-induced THz generation from the arrays of metallic particles was studied.

**Analytical model.** The basic object we will consider is an asymmetrical metallic particle (see Fig. 1). To simplify the model we will restrict ourselves to 2D-geometry assuming that the particle length along the y-axis is very large and that the optical and THz wavevectors $k$ are mostly in $xz$ plane. By this example we are going to demonstrate that even a flat incidence of the optical pulse on the array of particles like in Fig. 1 should produce a strong single-period THz pulse because of the inhomogeneous heat redistribution into the electronic subsystem.

Femtosecond laser pulse acting on the array of particles heats electrons near the surface (in the optical skin-layer). Here we assume nondestructive irradiation which means that electron thermal energy is lower than ~ 1 eV. To illustrate the process of temperature gradient formation we may consider a particle which has a maximal size $d$ along the $x$ axis of about 2-3 skin-depth and a bigger size $L$ along the z axis (see Fig. 1). In this case the duration of heat distribution along the $x$ axis is much lower than along the $z$ axis.

We will describe the electron heating and redistribution of the electron thermal energy $\varepsilon(x,z,t)$ with an equation of diffusion type:

$$\frac{\partial \varepsilon}{\partial t} = D\left(\frac{\partial^2 \varepsilon}{\partial x^2} + \frac{\partial^2 \varepsilon}{\partial z^2}\right) + \frac{\nu E^2 e^2}{2m\omega^2}, \quad (1)$$

where $E$ is the optical electric field, $\omega$ is the optical frequency, $\nu$ is the effective electron scattering rate, $m$ and $-e$ are free electron mass and charge respectively, $D$ is the thermal diffusivity which can be estimated as $D = v_F^2/(3\nu_e)$ ($v_F$ is the Fermi velocity and $\nu_e$ is the electron collision frequency which is in general case not equal to $\nu$). Note that in the case of metals electronic temperature $T$ and average energy $\varepsilon$ are in the following relation $\varepsilon \cong 3\varepsilon_F/5 + \pi^2 T^2/(4\varepsilon_F)$ [21]. From the eq. (1) we obtain the characteristic time of heat diffusion $\tau = l^2/D$, where $l$ is the characteristic spatial scale of the initial heat distribution. To estimate the value of $\tau$, $l$ should be chosen equal to the electron mean free path or to the skin-layer depth. On the boundary of metallic particle we assume zero thermal energy flux, thus boundary conditions are $\partial \varepsilon / \partial \mathbf{n}|_\Sigma = 0$, where $\mathbf{n}$ is the normal vector of particle surface $\Sigma$.

After the time $\tau$ electrons in the right side of the particle have homogeneous temperature over the entire depth, while in the left side heat redistribution is not yet over and the temperature is decreasing (see Fig. 1). The rough estimation gives the following maximal value of the thermal energy gradient along the $z$ axis:

$$\nabla \varepsilon \cong \frac{\delta \varepsilon}{L}, \quad (2)$$

where $\delta \varepsilon$ is an average thermal energy received by an electron near the surface due to the laser pulse action. The gradient will disappear due to longitudinal heat transport or due to the thermalization of electrons and the crystal lattice, which takes 1 – 10 ps depending on the metal type [4]. Note that the $\nabla \varepsilon$ estimation does not account for the electric field enhancement near the particle corners (this effect leads to an increase of the gradient (2) and will be discussed in the next section).

To describe the low-frequency radiation produced by the time-dependent thermal energy gradient we use a hydrodynamic approach developed in [10]. The basic equations are:

$$\frac{\partial \mathbf{v}}{\partial t} = -\frac{1}{mn}\nabla p - \frac{e\mathbf{E}}{m} - \nu \mathbf{v}, \quad (3)$$

$$\frac{\partial n}{\partial t} = -div\, n\mathbf{v}, \quad (4)$$

$$p = \frac{2}{3}n\varepsilon, \quad (5)$$

where $\mathbf{v}$ is the hydrodynamic electrons velocity, $p$ is the electron gas pressure, $n$ is the density of electrons (in cm$^{-3}$), $\mathbf{E}$ is the self-consistent electric field, $\nu$ is the effective collision frequency. Electric and magnetic fields are described by the Maxwell equations:

$$\text{rot}\, \mathbf{H} = \frac{1}{c}\frac{\partial \mathbf{E}}{\partial t} - \frac{4\pi}{c}ne\mathbf{v}, \quad (6)$$

$$\text{rot}\, \mathbf{E} = -\frac{1}{c}\frac{\partial \mathbf{H}}{\partial t}. \quad (7)$$

Here $\mathbf{H}$ is the magnetic field, $c$ is the speed of light in vacuum. By assuming ideal reflection of electrons from particle surface we can derive zero boundary condition for the normal electron velocity:

$$\mathbf{v}\,||\,\mathbf{n}|_\Sigma = 0. \quad (8)$$

No additional boundary conditions is needed for electromagnetic field, moreover with some ground electron energy $3\varepsilon_F/5$ normal electric field remains continuous at metallic particle boundary, having considerable variation inside Debye layer which can be derived from total electron energy $\varepsilon$.

From the system (4)-(6) and (9) we obtain (see [10] for details) the following equation for the electron density perturbation:

$$\frac{\partial^2 \delta n}{\partial t^2} + \nu \frac{\partial \delta n}{\partial t} + \omega_p^2 \delta n - \frac{2\varepsilon_F}{5m}\Delta \delta n = \frac{2n_0}{3m}\Delta \varepsilon, \quad (9)$$

where $\delta n = n - n_0$ is the electron density perturbation, $n_0$ is the initial electron density, $\omega_p = (4\pi n_0 e^2/m)^{1/2}$ is the Langmuir plasma frequency in metal, $\varepsilon_F$ is the Fermi energy, $\Delta$ is the Laplace operator. Taking into account that $\omega_p$ in metals is much higher than $\nu$, we find the value of $\delta n$ out of the Debye (Tomas-Fermi) layer:

$$\delta n = \frac{1}{6\pi e^2}\Delta \varepsilon. \quad (10)$$

Using the same assumptions we obtain an analytical expression for the electric field inside the particle from eq. (11) and Maxwell equations:

$$\mathbf{E} = -\frac{2}{3e}\nabla \varepsilon. \quad (11)$$

The above expression is valid inside the particle. Being also valid for tangential field in the vicinity of the most of the triangle surface it does not make proper estimation for radiated field, since the strong counter directed electric field appearing between particles.

The proper evaluation for radiated field can be achieved by solving the equation (12) for each time $t$ i.e. find a solution $\mathbf{E}_s$ for electrostatic problem with the given external source $\nabla \varepsilon (t)$. Then from the charge density $\rho_s(t) = \text{div}\,\mathbf{E}_s / 4\pi$ the planar dipole momentum along $z$ axis can be calculated:

$$d_s(t) = \int_{particle} \rho_s \, z \, dz \, dx. \quad (12)$$

The radiation magnetic field $B_y$ rad can be obtained from the slow varying dipole momentum as:

$$B_{y\,\text{rad}} = \frac{1}{2} \frac{1}{L_{\text{cell}}} \frac{\partial d_s}{\partial t}, \quad (13)$$

where $L_{\text{cell}}$ is the period of particles along $z$ direction. We do not account for dipole momentum along $x$ since are assuming quite long array of particles in the $z$ axis direction, so $x$- directed dipoles will not radiate in the $x$ direction. This quasi-static approach is suitable for comparatively shallow particles. When the particle depth (the scale along $x$) is greater or equal to the radiated wavelength the above formulae is not accurate enough (see the difference in the pictures Fig. 2 $a, b, c$). From the above it follows that the particle array works like a dipole antenna excited by the electron energy gradient. Note that if the particles are electrically connected no dipole moment is induced, thus no radiation occurs.

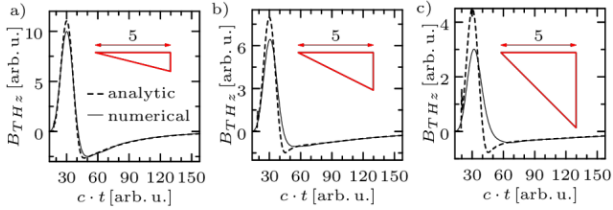

Fig. 2. The accuracy (better to worse) of quasi-static approach to radiation problem of sub-wavelength particle array. Solid curves are the result of numerical simulation of the full system (1), (3)-(7), dashed curves are obtained by substitution of $\varepsilon$ calculated from (1), (3)-(7) into (12), (13). The units of time $t$ in calculations were the same as coordinates units, particle sketches are given with true aspect.

**Numerical scheme and modeling results.** To prove the concept of THz generation from asymmetric metallic particles we used numerical simulations of the dynamic system described by equations (1) and (3)–(7). We set all the fields on shifted lattices similar to Yee lattice used for Maxwell equations. Since only $p$-polarized waves are considered, two components of the electric field and one component of the magnetic field were set. Electron velocity components were set in the same nodes as corresponding components of the electric field. The electron density and energy were set in same individual nodes to achieve best approximation of the differential operators in eqns. (3) and (4). Setting values of the fields at $t$ and $t + dt / 2$ time points allows to achieve second order approximation of the step operator advancing the system by time interval $dt$. For all the differential operators standard approximation with a minimal number of points was used.

All the time the initial conditions of the Cauchy problem were set to zero, while the femtosecond optical pulse was modeled with an external source of $E_z$ component set at some surface above the particle array. We used moving window concept in the case of oblique incidence. To eliminate the impact of the bounding box we used PML approach, which was fine-tuned for minimal reflection of the waves existing in the system. The numerical lattice used in this research is equidistant with the cell having 1:1 scale ratio in the $xz$ plane. This makes impossible to explicitly set triangle particle hypotenuse face, so all the geometry is Manhattan. This leads to some simplifications of boundary conditions.

Since the particle surface contour coincide with cells, describing the plasma density/energy fields, all the currents appear to be normal with respect to these cells boundary by the above design, so all the fields that are set on the particle surface contour can be treated as zeros in all the calculations. The energy fields continued as follows: instead of missing (out of the particle contour) nodes needed for standard five point scheme for Laplace operator in Eq. (1), the values from the central node can be taken. This effectively provides Von Neumann boundary conditions on the particle contour. More details regarding the lattice and the field values bindings can be found in the Appendix of [16].

The system of eqns. (1) and (3)–(7) can be rewritten in the dimensionless form except one parameter. We have chosen the thermal diffusivity $D$ to be such a parameter. Numerical step $dt$ was chosen according to Courant–Friedrichs–Lewy condition which is minimum between one from Maxwell set of equations $1 / dx$ and from the thermal conductivity equation $D / dx^2$. Typical parameters for numerical simulations were $\omega_{\text{opt}} / \omega_p = 0.14$, $D = 0.3\,\omega_p\,l_{\text{skin}}^2$, $L_p = 30\,l_{\text{skin}}$, where $l_{\text{skin}}$ is the optical skin depth, $\omega_p$ is Langmuir plasma oscillation frequency, $\omega_{\text{opt}}$ is femtosecond pulse frequency, $L_p$ is particle size along $z$ coordinate, corresponding value for $x$ coordinate was varied around $L_p$ value. Femtosecond pulse had 8 oscillations which is close to the typical value in laser induced THz experiments yet a bit lower to save computational resources. The laser pulse amplitude was set as $(1 - \cos(2\pi t / T)) \cdot \sin(2\pi n_{\text{osc}} t / T)$, where $T$ is the laser pulse duration and $n_{\text{osc}}$ is the integer number of full waves in the wave packet, so $\omega_{\text{opt}} = n_{\text{osc}}\,2\pi / T$.

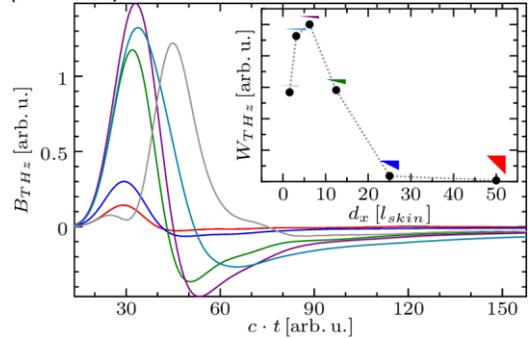

Fig. 3. THz signal generated from the nanoparticle array with different particle depths (normal incidence case). Inset: the dependence of the THz pulse full energy on the particle depth.

In the numerical modeling we found that for our parameters the optimal particle depth is about 5-6 times larger than the optical skin depth (see Fig. 3). At 800 nm wavelength it corresponds to the average particle thickness of 40-45 nm for silver or gold. The theoretical dependence of the THz pulse energy on the particle depth (see Inset in Fig. 3) is in good agreement with the experimental results on the THz generation from silver nanoparticles [18], despite the particle shapes were not the same as in our modeling. Moreover, the similar dependence was measured in the experiments on the optical-to-THz conversion on gold-coated gratings of different thickness [22] – the optimal coating depth was found to be of about 40 nm.

The dependencies of the THz waveform and full energy on the incidence angle of the laser pulse (see Fig. 4) were also analyzed numerically. As expected, the optical-to-THz conversion on the directed particle array is a strongly asymmetrical process: a half-turn of the sample around its normal leads to change of the THz pulse energy by 1-3 orders of magnitude (see Fig. 4, a). At the same time, the waveform is also strongly dependent on the angle and, in particular, can be almost reversed for two symmetric incidence directions (Fig. 4, b)).

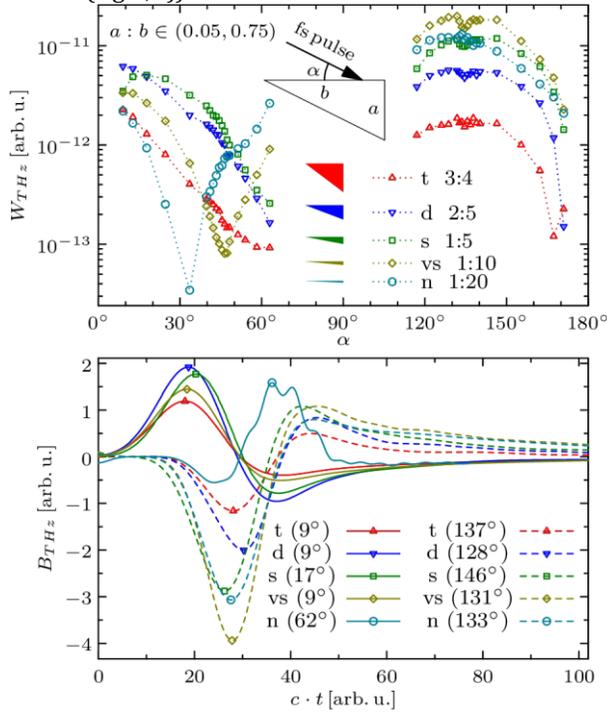

Fig. 4. THz generation efficiency depending on particle height (oblique incidence case) – top plot. THz waveforms for laser pulse incident from left and right sides at angles corresponding to maxima of generation efficiency shown in the top plot – bottom plot. The FWHM of the pump laser pulse is 36 arb. units.

We suppose that in the case of oblique incidence the superposition of two effects is observed: (1) the temperature gradient described in the previous section (directed toward the thin part of each particle); (2) the long-scale temperature gradient co-directed with the laser spot movement along the array. The first mechanism is a characteristic feature of asymmetrically heated particles only, while the second mechanism works also on a flat metal surface (see [10] for details). Depending on which side the laser pulse incidence from, these two effects can be of the same or of the opposite sign. Note that the obtained angular dependence of the THz energy (Fig. 4, a) is similar to the experimental one obtained for monocrystalline bismuth [23]. In this work a very strong asymmetry of the optical-to-THz conversion was found when the laser pulse incident obliquely on the sample face normal to the $C_1$ crystallographic axis.

**Conclusion.** Analytical and numerical study demonstrates that coherent THz generation from metal particles can be a consequence of the ultrafast laser heating of the electrons. We find that heat redistribution inside each nanoparticle leads to the temperature gradient formation and, further, to the generation of low-frequency polarization. Efficiency of optical-to-THz conversion in some of considered particle arrays estimated to be several times higher than in the case of the flat metal surface.

We observed a strong dependence of the calculated THz signal energy on the nanoparticles' shape, especially on their depth (see Fig. 3 and Fig. 4). The optimal average depth of the triangle particles found to be 2.5-3 times larger than the optical skin-layer thickness. It is comparable with the experimental results on the THz generation from arrays of triangle metal nanoparticles [18] and from metal gratings [13, 22]. Note that in this Letter we don't consider surface plasmon excitation which can be a possible mechanism of heating enhancement.

In the numerical modeling asymmetric generation from the directed nanoparticle pattern was also observed when the laser pulse incidence angle was varied (see Fig. 4). This should be treated as a proof-of-principle result showing that the thermal mechanism of optical-to-terahertz conversion allows to interpret strongly anisotropic THz response, like one observed earlier on the mono- and polycrystalline bismuth samples [23].

**Funding.** Russian Foundation for Basic Research (RFBR), grants No. 16-32-00717, 16-02-01078 and 17-02-00387.